\documentclass[journal]{IEEEtran}
\ifCLASSINFOpdf
\else
\fi
\usepackage{cite}
\usepackage{amsmath,amssymb,amsfonts}
\usepackage{algorithmic}
\usepackage{graphicx}
\usepackage{subcaption}
\usepackage{textcomp}
\usepackage{wrapfig}
\usepackage{algorithm}
\usepackage{lipsum}
\usepackage[hidelinks]{hyperref}
\usepackage{xcolor}
\usepackage{tabularx,booktabs}
\newcolumntype{C}{>{\centering\arraybackslash}X} 
\def\BibTeX{{\rm B\kern-.05em{\sc i\kern-.025em b}\kern-.08em
    T\kern-.1667em\lower.7ex\hbox{E}\kern-.125emX}}
\begin{document}
%
\title{A 2-$\mu$J, 12-class, $91\%$ Accuracy Spiking Neural Network Approach For Radar Gesture Recognition} 
%
%
%

\author{Ali Safa, \IEEEmembership{Student Member, IEEE}, Andr\'e Bourdoux, \IEEEmembership{Senior Member, IEEE}, Ilja Ocket, \IEEEmembership{Member, IEEE}, 
Francky Catthoor, \IEEEmembership{Fellow, IEEE}, 
Georges G.E. Gielen, \IEEEmembership{Fellow, IEEE}

\thanks{Ali Safa, Ilja Ocket, Francky Catthoor and Georges G.E Gielen are with imec and the Department of Electrical Engineering, KU Leuven, 3001 Leuven, Belgium (e-mail: Ali.Safa@imec.be; Ilja.Ocket@imec.be; Francky.Catthoor@imec.be; Georges.Gielen@kuleuven.be).}
\thanks{Andr\'e Bourdoux is with imec, 3001 Leuven, Belgium (e-mail: Andre.Bourdoux@imec.be).}
}
%
%

\markboth{}%
{Safa \MakeLowercase{\textit{et al.}}: }
%



\maketitle

\begin{abstract}
Radar processing via spiking neural networks (SNNs) has recently emerged as a solution in the field of ultra-low-power wireless human-computer interaction. Compared to traditional energy- and area-hungry deep learning methods, SNNs are significantly more energy efficient and can be deployed in the growing number of compact SNN accelerator chips, making them a better solution for ubiquitous IoT applications. We propose a novel SNN strategy for radar gesture recognition, achieving more than 91\% of accuracy on two different radar datasets. Our work significantly differs from previous approaches as 1) we use a novel radar-SNN training strategy, 2) we use quantized weights, enabling power-efficient implementation in real-world SNN hardware, and 3) we report the SNN energy consumption per classification, clearly demonstrating the real-world feasibility and power savings induced by SNN-based radar processing. We release evaluation code to help future research. 
\end{abstract}


%
\IEEEpeerreviewmaketitle

\section*{Supplementary Material}
Evaluation code is available at: \url{https://tinyurl.com/yu598c7e}
\section{Introduction}
\label{sec:introduction}
\IEEEPARstart{W}{ireless} human-computer interaction using radar-based gesture recognition systems has attracted large interest during the past decade, enabling applications such as smart domotics, AR/VR headsets and many other \textit{touchless} interfacing solutions that are key for a more hygienic, post-COVID-19 world \cite{lancet}. In order to embed radar sensing into ubiquitous, ultra-low-power IoT devices, research at the hardware side has mainly been devoted to high-level integration of radar transceivers \cite{intergration} with a focus on energy and area efficiency \cite{ourradar}. In contrast, research at the signal processing side has mainly been devoted to the use of high-accuracy \textit{deep neural networks} (DNNs) known to be rather energy- and area-hungry \cite{b12, b13}. State-of-the-art DNN-based techniques either rely on the use of an expensive desktop-grade GPU \cite{b12} or either on the use of a lower-power and lower-area embedded GPU (e.g., 10-W Nvidia Jetson Nano) \cite{b13}, still ill-suited for ultra-low-power applications like ubiquitous IoT.

\begin{figure}[!t]
\centering
    \includegraphics[scale = 0.6]{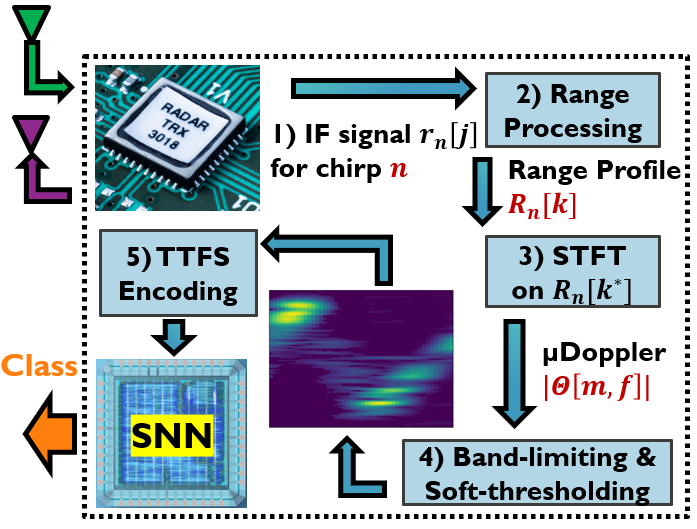}
    \caption{\textit{\textbf{$\boldsymbol{\mu}$Doppler-based radar-SNN architecture} proposed to solve the 5-class 8-GHz dataset of \cite{ubrain} with $93\%$ of accuracy}}
    \label{blockdia}
\end{figure}

\begin{figure}[!t]
\centering
    \includegraphics[scale = 0.65]{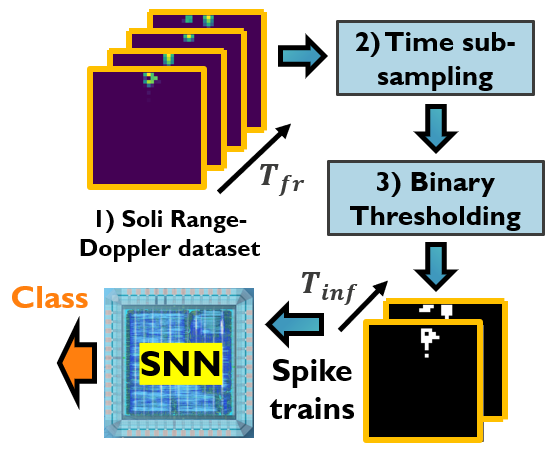}
    \caption{\textit{\textbf{Range-Doppler radar-SNN architecture} proposed to solve the \textbf{12-class} dataset of \cite{b12} with $91\%$ of accuracy.}}
    \label{blockdia2}
\end{figure}

Very recently, the use of energy-efficient \textit{spiking neural networks} (SNNs) for radar processing has grown to become an emerging topic in radar sensing and is currently being investigated by many teams \cite{ubrain,inton,infineon,b11}. Algorithm-wise, SNNs differ from DNNs as they communicate inter-neural information asynchronously, using binary spikes that are only emitted when the neuron membrane potential reaches a specific threshold. In contrast to DNNs, SNNs do not require expensive \textit{multiply-accumulate} operations at the input of each neuron, but make use of inexpensive \textit{add} operations only. Hardware-wise, SNNs can be integrated near the radar sensor (see Fig. \ref{blockdia}) as \textit{sub-threshold} analog circuits, reaching more than 5 orders of magnitude lower power consumption compared to embedded GPUs \cite{b13,analog,loihi}. 

Still, the development of SNN-based radar processing is at an early stage. In this letter, our aim is to propose a novel SNN architecture for radar gesture recognition using a different approach than the ones used in previously presented radar-SNN systems. Compared to previous works \cite{ubrain,infineon,b11}, which either use the $\mu$Doppler pre-processing \cite{b18} or the range-Doppler pre-processing \cite{b12}, we demonstrate that our novel radar-SNN approach is compatible with both pre-processing techniques. In contrast to the work in \cite{inton}, our approach is purely SNN-based, while the system of \cite{inton} uses an SNN followed by classical machine learning techniques such as \textit{Random Forest}, which cannot be deployed in sub-threshold analog SNN circuits. Compared to \cite{inton,infineon,b11}, our system uses implementation-ready, quantized weights (typical bit width in SNN hardware is $<8$ bits \cite{odin,ubrain}), while none of the aforementioned works quantize their weights, making their reported performances (85\%-98\%) \textit{unclear} when deployed in real-world hardware (12-class $91\%$ with 6-bit weights and 5-class $93\%$ with 4-bit weights in our work). Finally, in contrast to most previously mentioned works \cite{inton,infineon,b11}, we report an estimate of our SNN energy consumption when deployed in dedicated SNN hardware \cite{ubrain}. We assess the performance of our system on two different radar gesture datasets: the 12-class Google Soli dataset of \cite{b12} and the 5-class 8-GHz dataset of \cite{ubrain}. We report a $\times 3$ increase in the number of gesture classes compared to state-of-the-art quantized-weight SNNs.

\begin{figure*}[!t]
\centering
\captionsetup{justification=centering}
    \includegraphics[scale = 0.58]{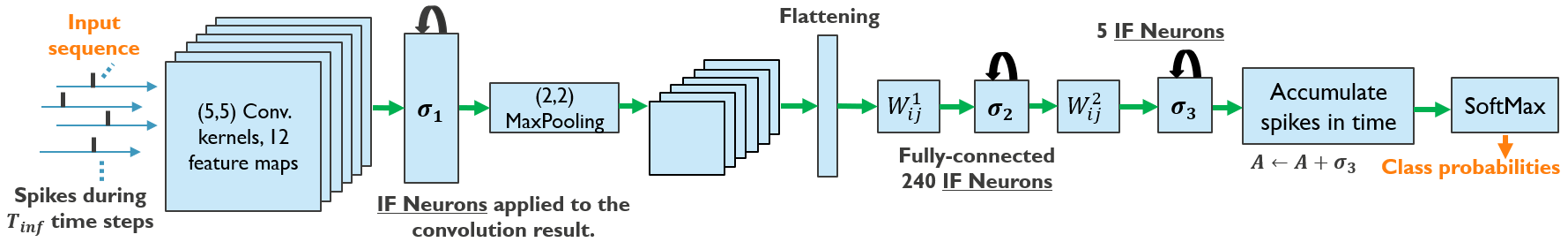}
    \caption{\textit{\textbf{SNN architecture used for radar processing}. Each spiking map slice corresponding to each time step is fed one by one to the network and the IF neurons change state according to their self recurrence (as denoted by the black recurrence arrows). 
    }}
    \label{fig1}
\end{figure*}

\section{Radar-SNN Processing Pipeline}
\subsection{5-class 8-GHz dataset and pre-processing}
The dataset of \cite{ubrain} contains radar ADC data with $N_{chirps}=192$ chirps per frame and with a variable number of frames per gesture acquisition $N_{frames}$ (step 1 in Fig. \ref{blockdia}). $\mu$Doppler signatures \cite{b18} are acquired for each gesture acquisition in the dataset by first computing the range profiles $R_n[k]$ for each chirp $n = 1,...,N_{tot}$ (where $N_{tot}$ is the total number of chirps). $R_n[k]$ is acquired by DFT using a Blackman window \cite{harris} (step 2 in Fig. \ref{blockdia}). Then, we apply the \textit{Short-Time Fourier Transform} (STFT) to the sequence $\Tilde{R}_n[k^*] = R_{n}[k^*] - R_{n-1}[k^*]$ (step 3 in Fig. \ref{blockdia}), which removes the strong DC component during each analysis window \cite{vars}, as follows:
\begin{equation}
    \Theta[m,f] = \sum_{n = -\infty}^{\infty} \Tilde{R}_n[k^*] g_{s}[n-mR] e^{-j2\pi fn}
    \label{eq10}
\end{equation}
where $k^*$ denotes the range bin where the gestures are executed, $g_s$ denotes a \textit{Hanning} window of length $s$, 
and $R$ is the hop size ($s = 192$ and $R=8$ throughout this paper). $k^*$ is known \textit{a priori} as the gestures are executed at 2 meters from the radar. We define the $\mu$Doppler signature as $|\Theta[m,f]|$ which is a matrix of size $(N_T \times s)$ with $N_T$ given by \cite{b11}:
\begin{equation}
    N_T = \left \lfloor{\frac{N_{frames} N_{chirps} - N_{overlap}}{R}}\right \rfloor
    \label{count}
\end{equation}
where $N_{overlap} = s - R$ is the number of overlapping bins between successive windows. Radar maps to be fed to the SNN are obtained by cutting $|\Theta[m,f]|$ along dimension $m$ into images of $48$ time samples. $\left \lfloor{\frac{N_T}{48}}\right \rfloor$ examples are thus obtained for each acquisition. By balancing the dataset and by removing the first and the last $6$ example maps to remove start-up (when the human simply sits in front of the radar before performing gestures) and ending artefacts (when the human reaches out to the radar to stop it), we obtain a balanced dataset with a total of $1695$ $\mu$Doppler examples.

Each example map is then normalized between $[0, 1]$. Out-of-band noise is removed through the band-limiting of the Doppler frequency axis by keeping the normalized frequency range between $[-0.26, 0.26]$ only. This frequency band was identified visually by evaluating the maximal significant extent of the Doppler spectra in the dataset. Then, we use \textit{soft thresholding} \cite{b19} to remove in-band noise in each Doppler spectrum (step 4 in Fig. \ref{blockdia}). The soft thresholding is performed by keeping the $k$ largest values and pad the remaining ones to $0$. We choose $k$ heuristically by considering that more than half of the Doppler samples within the normalized frequencies $[-0.26, 0.26]$ are significant (i.e. not noise), which leads to the choice of $k= \left \lfloor{\frac{192 \times (0.26 - (-0.26))}{2}}\right \rfloor - 1 = 48$ (we tried other $k$ values around $48$, but did not observe any significant boost in SNN accuracy). After step 4, Fig. \ref{blockdia} shows an example radar map, resulting from the $\mu$Doppler pre-processing described above.

The pre-processed radar $\mu$Doppler maps must then be converted into event streams to be compatible with the spiking nature of our SNN. Each pixel of the map is coded as a spike train of length $T_{inf}$ (number of time steps per inference). We encode each pixel using \textit{Time-To-First-Spike} (TTFS) encoding (step 5 in Fig. \ref{blockdia}), where a pixel of value $v \in [0,1]$ is quantized into an event train containing one spike located at index $T_{inf}-\left \lfloor{v T_{inf}}\right \rfloor$ \cite{ttfs}. If the pixel is equal to 0, then no spikes are emitted. As we are aiming at low-latency inference, we choose $T_{inf} = 4$ time steps.
\subsection{12-class Soli dataset and pre-processing}
The Soli dataset \cite{b12} has been acquired using a 60-GHz FMCW radar and is composed of 12 classes with a total of 5500 CFAR-processed range-Doppler magnitude acquisitions. Each gesture acquisition is a collection of maps $RD[t,l,m]$ where $t$ is the frame index, $l$ is the range index and $m$ is the Doppler index (see step 1 in Fig. \ref{blockdia2}), with a varying number of time steps $t \in [1,T_{fr}]$ per acquisition. First, we average and sub-sample each gesture acquisition $RD[t,l,m]$ (with varying $T_{fr}$) along $t$ (step 2 in Fig. \ref{blockdia2}) to a fixed number $T_{inf} < T_{fr}$ $\forall T_{fr}$ of frames per acquisition, as follows:
\begin{equation}
    RD[n,l,m] = \frac{T_{inf}}{T_{fr}} \sum_{t=n}^{n+\frac{T_{fr}}{T_{inf}}} RD[t,l,m]
    \label{eqavrg}
\end{equation}
where $n$ is the sub-sampled time index. Then, the resulting frames are converted to binary images $RD_b[n,l,m]$ by thresholding against $0$ (step 3 in Fig. \ref{blockdia2}). Therefore, for any pixel coordinate $(l^*,m^*)$, $RD_b[n,l^*,m^*]$ represents a spike train of length $T_{inf}$, set to $28$ (minimum $T_{fr}$ in the dataset). 

\subsection{Spiking neural network for classification}
\label{sec:arch}

To classify the spiking radar tensors, we use the SNN architecture shown in Fig. \ref{fig1} with \textit{Integrate and Fire} (IF) neurons:
\begin{equation}
\begin{cases}  V^{k+1} = V^{k} + J_{in} \hspace{5pt} \mbox{and} \hspace{5pt} S = 0 & \mbox{if } V^{k} < 1 \\ V^{k+1} = 0 \hspace{5pt} \mbox{and} \hspace{5pt} S = 1 & \mbox{if } V^{k} \geq 1 \end{cases}
\label{eq13}
\end{equation}
where $V^k \geq 0$ is the neural membrane potential at time step $k$, $J_{in}$ is the neuron input and $S$ is the spiking output. As the derivative of spikes as a function of the membrane potential is ill-defined, we create a custom neuron model using the \textit{pyTorch} framework \cite{pytorch} which behaves as (\ref{eq13}) in forward pass. For the backward pass, we approximate the derivative using a Gaussian function (\ref{eq14}) as the surrogate derivative \cite{b5}. This enables the use of back-propagation in the spiking domain.
\begin{equation}
    \sigma'(V) \approx \frac{1}{\sqrt{2 \pi}} e^{-2V^2}
    \label{eq14}
\end{equation}
The layer-by-layer description of our SNN architecture (Fig. \ref{fig1}) is the following. After the spike train encoding of the radar maps, we use a $(5,5,12)$ convolutional layer. At each time step, the convolution result is fed to the IF neuron layer $\sigma_1$. Then, the spiking tensor at the output of $\sigma_1$ is down-sampled via \textit{MaxPooling} and the resulting tensor is flattened to a 1-dimensional spiking vector. Then, two fully-connected spiking layers are used and the 12- or 5-dimensional output of $\sigma_3$ (corresponding to the 12 or 5 gesture classes) is accumulated over time in a vector $A$. Finally, $A$ is transformed via \textit{SoftMax} into class probabilities. Our network architecture search was conducted with the objective of achieving a $>90\%$ accuracy with heavily quantized weights (at most 6-bit) and a small network size.

For training, we use the \textit{Adam} optimizer \cite{b20} with learning rate $10^{-3}$. The batch size is $128$ and the SNN is first trained for 14 epochs with full-bit weights and 1 epoch with quantized weights in the forward pass and full-bit weights in the backward pass. The accuracy of our SNN is assessed using 6-fold cross validation. 

\section{Experimental Results}
\label{sec:expres}
Table \ref{datasetcontent} reports the performance of our proposed system (entry 6 for the 8-GHz dataset and entry 7 for the Soli dataset) against the state of the art. We evaluate the energy per classification $E_c$ of our SNN using the hardware metrics of the $\mu$Brain SNN chip, described in \cite{ubrain}:
\begin{equation}
    E_c = N_{spikes} \times E_{dyn} + \delta T \times P_{stat}
\end{equation}
where $N_{spikes}$ is the maximum number of spikes during classification, $E_{dyn}=2.1$ pJ is the energy per spike, $P_{stat} = 73$ $\mu$W is the static leakage power and $\delta T$ is the inference time. Even though a smaller $\delta T$ can be reached by adjusting the bias voltages that control the delay cells in \cite{ubrain}, we assume $\delta T=4$ ms for the 8-GHz dataset ($T_{inf}=4$) and $\delta T=28$ ms for the Soli dataset ($T_{inf}=28$) to provide an upper bound estimate on $E_c$.

\begin{table}[htbp]
\begin{center}
\begin{tabular}{|c|c|c|c|c|}
\hline
\textbf{Architecture} & $\boldsymbol{N_{c}}$ & \textbf{Accuracy}  & $\boldsymbol{E_c}$ & $\boldsymbol{N_{bits}}$  \\
\hline
\textbf{1)} DNN \cite{b13}  &  12 & 94\%  & 330 mJ & 32-f \\
\hline
\textbf{2)} SNN-STDP \cite{b11}  &  8 & 85\%  & - & 32-f \\
\hline
\textbf{3)} SNN-conv \cite{infineon}   &  4 &   98.5\%  & - & 32-f \\
\hline
\textbf{4)} SNN-RF \cite{inton} & 11 & 98\%  &  - & 4-i \& 32-f \\
\hline
\textbf{5)} SNN-conv \cite{ubrain}  &  4 & 93.4\%  & 340 nJ & 4-i \\
\hline
\textbf{6)} \textbf{This work} (8-GHz)  & 5 &  $93 \pm 2$\%   & 351 nJ & 4-i \\
\hline
\textbf{7)} \textbf{This work} (Soli)  & \textbf{12} &  $91 \pm 1$\%   &  2 $\mu$J & 6-i \\
\hline
\end{tabular}
\caption{\textit{\textbf{Our proposed system compared to the state of the art.} $N_c$ is the number of classes, $E_c$ is the energy consumption per classification (not reported for entries 2-4) and $N_{bits}$ is the number of bits for the network weights (f and i stand for float and integer respectively).}}
\label{datasetcontent}
\end{center}
\end{table}

Out of the implementation-ready SNNs using quantized weights \textit{only} (entries 5, 6 and 7 in Table \ref{datasetcontent}), our work significantly outperforms entry 5 by up to $\times 3$ higher number of gesture classes $N_c$, while having a similar accuracy and $N_{bits}$, with $E_c$ of the same order of magnitude. All other entries in Table \ref{datasetcontent} either rely on DNNs (entry 1) and RF (entry 4), being ill-suited for ultra-low-power IoT, or do not quantize their weights (entries 2 and 3, giving unclear performance in real-world SNN hardware). In addition, our work achieves a recognition accuracy close to the accuracy of the DNN in entry 1 \cite{b13}, while consuming more than two orders of magnitude less energy per inference. Finally, entries 6 and 7 clearly show how our system trades off $N_c$, $E_c$ and $N_{bits}$ for a target accuracy of $>90\%$.      

\section{Conclusion}
\label{sec:conc}
This letter has presented a novel radar-SNN architecture for ultra-low-power radar gesture recognition, significantly outperforming existing implementation-ready SNNs in terms of classification performance. The presented approach has reported several key innovations compared to previous radar-SNN systems such as a novel radar-SNN training strategy and radar to spike encoding approaches. Radar-SNN evaluation code has also been provided, which helps lighting the way for the emerging area of SNN-based radar processing. 

\section*{Acknowledgment}
\label{sec:ack}
The authors thank Dr. Federico Corradi and Dr. Lars Keuninckx for the discussions and guidance, and the Flanders AI research program for partially supporting this work.


\begin{thebibliography}{00}

\bibitem{lancet} Chin, A., Chu, J., Perera, M., Hui, K., Yen, H. L., Chan, M., Peiris, M., Poon, L. (2020). "Stability of SARS-CoV-2 in different environmental conditions." The Lancet. Microbe, 1(1), e10. https://doi.org/10.1016/S2666-5247(20)30003-3

\bibitem{intergration} J. Rimmelspacher, R. Ciocoveanu, G. Steffan, M. Bassi and V. Issakov, "Low Power Low Phase Noise 60 GHz Multichannel Transceiver in 28 nm CMOS for Radar Applications," 2020 IEEE Radio Frequency Integrated Circuits Symposium (RFIC), 2020, pp. 19-22, doi: 10.1109/RFIC49505.2020.9218297.

\bibitem{ourradar} Y. Liu et al., "9.3 A680 $\mu$W Burst-Chirp UWB Radar Transceiver for Vital Signs and Occupancy Sensing up to 15m Distance," 2019 IEEE International Solid- State Circuits Conference - (ISSCC), San Francisco, CA, USA, 2019, pp. 166-168, doi: 10.1109/ISSCC.2019.8662536.


\bibitem{b12} Wang, S., Song, J., Lien, J., Poupyrev, I., Hilliges, O. (2016). "Interacting with Soli: Exploring Fine-Grained Dynamic Gesture Recognition in the Radio-Frequency Spectrum," (pp. 851-860).

\bibitem{b13} Y. Sun, T. Fei, X. Li, A. Warnecke, E. Warsitz and N. Pohl, "Real-Time Radar-Based Gesture Detection and Recognition Built in an Edge-Computing Platform," in IEEE Sensors Journal, vol. 20, no. 18, pp. 10706-10716, 15 Sept.15, 2020, doi: 10.1109/JSEN.2020.2994292.

\bibitem{ubrain} Stuijt, J., Sifalakis, M., Yousefzadeh, A., Corradi, F. (2021). "$\mu$Brain: An Event-Driven and Fully Synthesizable Architecture for Spiking Neural Networks." Frontiers in Neuroscience, 15, 538.

\bibitem{inton} Tsang, I. et al., (2021). "Radar-Based Hand Gesture Recognition Using Spiking Neural Networks." Electronics, 10(12).

\bibitem{infineon} Arsalan M. et al., (2021). "Resource Efficient Gesture Sensing Based on FMCW Radar using Spiking Neural Networks"

\bibitem{b11} D. Banerjee et al., "Application of Spiking Neural Networks for Action Recognition from Radar Data," 2020 International Joint Conference on Neural Networks (IJCNN), Glasgow, United Kingdom, 2020, pp. 1-10, doi: 10.1109/IJCNN48605.2020.9206853.

\bibitem{odin}C. Frenkel, M. Lefebvre, J. Legat and D. Bol, "A 0.086-mm$^2$ 12.7-pJ/SOP 64k-Synapse 256-Neuron Online-Learning Digital Spiking Neuromorphic Processor in 28-nm CMOS," in IEEE Transactions on Biomedical Circuits and Systems, vol. 13, no. 1, pp. 145-158, Feb. 2019, doi: 10.1109/TBCAS.2018.2880425.

\bibitem{analog} S. Moradi, N. Qiao, F. Stefanini and G. Indiveri, "A Scalable Multicore Architecture With Heterogeneous Memory Structures for Dynamic Neuromorphic Asynchronous Processors (DYNAPs)," in IEEE Transactions on Biomedical Circuits and Systems, vol. 12, no. 1, pp. 106-122, Feb. 2018, doi: 10.1109/TBCAS.2017.2759700.

\bibitem{loihi} M. Davies et al., "Loihi: A Neuromorphic Manycore Processor with On-Chip Learning," in IEEE Micro, vol. 38, no. 1, pp. 82-99, January/February 2018, doi: 10.1109/MM.2018.112130359.

\bibitem{b18} Victor C. Chen (2014). "Radar Micro-Doppler Signatures: Processing and Applications," Institution of Engineering and Technology.

\bibitem{harris} F. J. Harris, "On the use of windows for harmonic analysis with the discrete Fourier transform," in Proceedings of the IEEE, vol. 66, no. 1, pp. 51-83, Jan. 1978, doi: 10.1109/PROC.1978.10837.

\bibitem{vars} B. Vandersmissen, N. Knudde, A. Jalalvand, I. Couckuyt, A. Bourdoux, W. De Neve, and T. Dhaene, “Indoor person identification using a low-power fmcw radar,” IEEE Transactions on Geoscience and Remote Sensing, vol. 56, no. 7, pp. 3941–3952, July 2018

\bibitem{b19} H. Xu, Z. Wang, H. Yang, D. Liu and J. Liu, "Learning Simple Thresholded Features With Sparse Support Recovery," in IEEE Transactions on Circuits and Systems for Video Technology, vol. 30, no. 4, pp. 970-982, April 2020, doi: 10.1109/TCSVT.2019.2901713.

\bibitem{ttfs} B. Rueckauer and S. Liu, "Conversion of analog to spiking neural networks using sparse temporal coding," 2018 IEEE International Symposium on Circuits and Systems (ISCAS), Florence, 2018, pp. 1-5, doi: 10.1109/ISCAS.2018.8351295.

\bibitem{pytorch} Paszke, A., Gross, S., Chintala, S., Chanan, G., Yang, E., DeVito, Z., Lin, Z., Desmaison, A., Antiga, L., Lerer, A. (2017). Automatic differentiation in PyTorch.

\bibitem{b5} E. O. Neftci, H. Mostafa and F. Zenke, "Surrogate Gradient Learning in Spiking Neural Networks: Bringing the Power of Gradient-Based Optimization to Spiking Neural Networks," in IEEE Signal Processing Magazine, vol. 36, no. 6, pp. 51-63, Nov. 2019, doi: 10.1109/MSP.2019.2931595.

\bibitem{b20} Kingma, D., Ba, J. (2014). "Adam: A Method for Stochastic Optimization," International Conference on Learning Representations.

\end{thebibliography}
\end{document}